\documentstyle[12pt]{article}
\begin{document}

\begin{center}
{\large\bf  Electroweak moments of baryons and hidden strangeness of the nucleon.
\footnotemark\\}%
\addtocounter{footnote}{0}\footnotetext
{Supported in part by the RFBR under contract 96-02-18137}
\end{center}
\begin{center}
S.B.Gerasimov \\
{\em Bogoliubov Laboratory of Theoretical Physics,\\
Joint Institute for Nuclear Research, 141 980 Dubna, Moscow region, Russia}\\
\end{center}
\vspace{1cm}

\begin{abstract}
The phenomenological sum-rule-based approach is used to discuss the quark
composition dependence of some static electroweak characteristics
of baryons.The role of nonvalence degrees of freedom ( the sea
partons and/or peripheral meson currents ) is shown to be important to select
and make use of the relevant symmetry parametrization of baryon observables.
The implications of the baryon magnetic moment analysis for estimation
of the $\Delta$q, (q = u, d, s), values of the spin-dependent DIS on nucleons,
the contribution of hidden strangeness to the nucleon magnetic moment and to
the quark-line-rule violating $\phi \pi$-production in antinucleon-nucleon
annihilation reaction are presented.

\end{abstract}

\section {\bf Outline and summary of sum rule approach to $\mu(B)$}

\baselineskip 20pt
The rapidly growing precision of measurements of the electroweak coupling
constants, like magnetic moments or the $G_A/G_V$-- ratios in baryon
semi-leptonic decays \cite{a_1,a_2,a_3}, may be the basis of new, more subtle and detailed
information on the internal structure of baryons.
In this report we consider some consequences of sum rules for the static,
electroweak characteristics of baryons following from the theory of broken
internal symmetries and common features of the quark models
including relativistic effects and corrections due to
nonvalence degrees of freedom -- the sea partons and/or the meson clouds
at the periphery of baryons.\\
In Ref. \cite{a_4,a_5}, the following parametrization was introduced for
magnetic moments $\mu(B)$ of baryons :
\begin{eqnarray}
&& \mu(B)=\mu(q_e) g_2 + \mu(q_o) g_1 + C(B) + \Delta, \label{e1}\\
&& \mu(\Lambda)=\mu(s) ({2\over 3} g_2 -{1\over 3} g_1) + (\mu(u) + \mu(d))
   ({1\over 6} g_2 + {2\over 3} g_1) + \Delta, \label{e2}\\
&& \mu(\Lambda\Sigma)={1\over \sqrt{3}} (\mu(u) - \mu(d))
   ({1\over 2} g_2 - g_1) + C(\Lambda\Sigma), \label{e3}\\
&& \Delta = \sum\limits_{q=u,d,s} \mu(q) \delta(N),\label{e4}
\end{eqnarray}
where $\mu(q)$ are the effective quark magnetic moments defined without
any nonrelativistic approximations, $q_e=q_{even}=u;d;s$ for $P$ and
$\Sigma^{+}$; $N$ and $\Sigma^{-}$; $\Xi^{o}$ and $\Xi^{-}$,
$q_o=q_{odd}=u;d;s$ for $N$ and $\Xi^o$; $P$ and $\Xi^{-}$; $\Sigma^{+}$ and $\Sigma^{-}$,
respectively, so that $P=u^{2}d$, $N=d^2u$, etc. if only valence quarks are
retained, $g_{2(1)}$ are "reduced" dimensionless
coupling constants obeying exact $SU(3)$--symmetry and related with the SU(3)
$F-$ and $D-$ type constants via $g_2=2F$ and $g_1=F-D$, $\delta(B)$ is a
matrix element of the OZI--suppressed $\overline qq$--configuration for
a given hadron: $\delta(B)=<B|\overline qq |B>$, where $q\not\subset
\{q_e^2, q_o\}$, e.g. $\delta(N)=<N|\overline ss |N>$, etc.
For the models, explicitly including the meson degrees of freedom, we
introduce the exchange current contribution to $\mu(B)$ due only to
the dominant pion currents (these contributions are represented by
the diagrams with the photon touching the charged pion line that
connects different nonstrange quark lines). By inspection, this
contribution is absent for all octet, three-quark hyperons except for
the nondiagonal $\mu(\Lambda\Sigma)$--magnetic moment.
So that $C(P)=-C(N)$ and $C(\Lambda\Sigma)$ are the isovector contributions
of the charged--pion exchange current to $\mu(P), \mu(N)$ and the
$\Lambda\Sigma$--transition moment $\mu(\Lambda\Sigma)$.
In the following we make use, consecutively, two pictures of the baryon
internal composition.In the first one, all baryons are considered
as consisting of three massive, "dressed" constituent quarks, locally
coupled with lightest goldstonions -- the pion fields.In the second picture
only fundamental QCD quanta -- the quarks and gluons -- are there, the
meson component of the baryon state vectors being represented by the
properly correlated "current" quarks and gluons.The use of one picture or
another will be reflected in a particular parametrization of
contributions due to corresponding nonvalence degrees of freedom.
Below, we shall use the particle and quark symbols for corresponding
magnetic moments. Equations (\ref{e1}) -- (\ref{e4}) lead to the
following sum rules \cite{a_4,a_5}:
\begin{eqnarray}
&& P + N + \Xi^0 + \Xi^- -3\Lambda + {1\over 2} (\Sigma^+ + \Sigma^-) = 0, \label{e5}\\
&& (\Sigma^+ - \Sigma^-) (\Sigma^+ + \Sigma^- - P - N) - (\Xi^0 - \Xi^-)
    (\Xi^0 + \Xi^- - P - N) = 0, \label{e6}\\
&& \alpha ={D\over F+D} = {g_2 - 2g_1\over 2(g_2-g_1)} = {1\over 2}
   \left(1 - {\Xi^0 - \Xi^-\over \Sigma^+ - \Sigma^- -
   \Xi^0 + \Xi^-} \right),\label{e7}\\
&& C(P) ={1\over 2} (C(P)-C(N)) = {1\over 2} (P - N + \Xi^0 - \Xi^-
   - \Sigma^+ + \Sigma^-), \label{e8}\\
&& C(\Lambda\Sigma) = \mu(\Lambda\Sigma) + {1\over\sqrt{3}}
   (\Xi^0 - \Xi^- - {1\over 2} (\Sigma^+ - \Sigma^-)), \label{e9}\\
&& {u-d\over u-s} = {\Sigma^+ - \Sigma^- - \Xi^0 + \Xi^- \over
    \Sigma^+ - \Xi^0}, \label{e10}\\
&& {u\over d} ={\Sigma^+ (\Sigma^+ - \Sigma^-) - \Xi^0 (\Xi^0 - \Xi^-) \over
   \Sigma^- (\Sigma^+ - \Sigma^-) - \Xi^- (\Xi^0 - \Xi^-)}, \label{e11}\\
&& {s\over d} ={\Sigma^+ \Xi^- - \Sigma^-\Xi^0 \over
   \Sigma^- (\Sigma^+ - \Sigma^-) - \Xi^- (\Xi^0 - \Xi^-)}, \label{e12}\\
&& \Delta^{++} = {u\over s} \Omega^- = {u\over d} \Delta^{--}. \label{e13}
\end{eqnarray}
Reserving the possibility of $g_i(N)\not=g_i(Y), Y=\Lambda, \Sigma, \Xi$,
due to a more prominent role of the pion degrees of freedom in the nucleon
and combining Eqs. (\ref{e5})--(\ref{e6}), we propose also the following,
probably, most general sum rule of this approach
\begin{eqnarray}
&&(\Sigma^+ - \Sigma^-) (\Sigma^+ + \Sigma^- - 6\Lambda + 2\Xi^0 + 2\Xi^-)\nonumber\\
&&- (\Xi^0 - \Xi^-) (\Sigma^+ + \Sigma^- + 6\Lambda - 4\Xi^0 - 4\Xi^-) = 0.
\label{e14}\end{eqnarray}
 Eqs.(\ref{e11}) -- (\ref{e13}) are obtained provided $\delta(N)=0$.
Hence, we relate them to the chiral constituent
quark models, where we assume
that a given baryon consists of three "dressed"
constituent quarks, and also
to the validity of the OZI (or the quark--line) rule.
Now, we list some
consequences of the obtained sum rules. By definition, the $\Lambda$--value
entering into Eqs. (\ref{e5}) and (\ref{e14}) should be "refined" from the
electromagnetic $\Lambda\Sigma^0$--mixing affecting  $\mu(\Lambda)_{exp}$.
Hence, the numerical value of $\Lambda$, extracted from Eq. (\ref{e14}),
can be used to determine the $\Lambda\Sigma^0$--mixing angle through the
relation
\begin{eqnarray}
\sin \theta_{\Lambda\Sigma} \simeq \theta_{\Lambda\Sigma} =
{\Lambda -\Lambda_{exp} \over 2\mu(\Lambda\Sigma)} = (1.43 \pm 0.31) 10^{-2}
\label{e15}\end{eqnarray}
in accord with the independent estimate of $\theta_{\Lambda\Sigma}$ from
the electromagnetic mass-splitting sum rule \cite{a_6}. Equation (\ref{e11})
shows that owing to interaction of the $u$-- and $d$-- quarks with charged
pions the "magnetic anomaly" is developing, i.e. $u/d=-1.80\pm 0.02 \not=
Q_u/Q_d = -2.$ Evaluation of the lowest order quark--pion loop diagrams
gives \cite{a_7}: $u/d=(Q_u +\kappa_u)/(Q_d+\kappa_d)=-1.77$, where $\kappa_q$
is the quark anomalous magnetic moment in natural units.
The meson--baryon universality of the quark characteristics, Eqs.
(\ref{e10}) -- (\ref{e12}), suggested long ago \cite{a_8}, is confirmed by
the calculation of the ratio of $K^*$ radiative widths
\begin{eqnarray}
{\Gamma(K^{*+}\to K^+\gamma)\over \Gamma(K^{*+}\to K^+\gamma)} =
\left({u/d} + {s/d}\over 1+ {s/d} \right)^2 =0.42\pm 0.03 \
(\mbox{vs} \ 0.44\pm 0.06 \mbox{\cite{a_1}}).
\label{e16}\end{eqnarray}
The experimentally interesting quantities $\mu(\Delta^+ P) = \mu(\Delta^0 N)$
and $\mu(\Sigma^{*0}\Lambda)$ are affected by the exchange current
contributions and for their estimation we need additional assumptions.
We use the analogy with the one--pion--exchange current, well--known in
nuclear physics, to assume for the exchange magnetic moment operator
\begin{eqnarray}
\hat\mu_{exch} =\sum\limits_{i<j} [\vec\sigma_i \times \vec\sigma_j]_3
[\vec\tau_i \times \vec\tau_j]_3 f(r_{ij}),
\label{e17}\end{eqnarray}
where $f(r_{ij})$ is a unspecified function of the interquark distances,
$\vec\sigma_i (\vec\tau_i)$ are spin (isospin) operators of quarks.
Calculating the matrix elements of $\hat\mu_{exch}$ between the baryon wave
functions, belonging to the 56--plet of $SU(6)$, one can find
\begin{eqnarray}
&& C(P)={1\over \sqrt{2}} C(\Delta^+P) = \sqrt{3} C(\Lambda\Sigma),\label{e18}\\
&& \mu(\Delta^+P; \{56\}) = {1\over \sqrt{2}} \left(P - N +{1\over 3} (P+N)
   {1-u/d\over 1+ u/d}\right). \label{e19}
\end{eqnarray}
where Eq.(19) may serve as a generalization of the well--known
$SU(6)$--relation  \cite{a_9}.
We close this section presenting the set of sum rules following
from Eqs.(2)--(4) with C(B's) =$ \delta$(B's) = 0, u/d=-2:
\begin{eqnarray}
&& \Sigma^+ [\Sigma^-] = P[-P-N] +
(\Lambda - {N\over 2})( 1+{{2N}\over P}),\label{e20}\\
&& \Xi^0[\Xi^-] = N[-P-N] + 2(\Lambda -
{N\over 2})( 1+ {N\over {2P}}),\label{e21}\\
&& \mu(\Lambda\Sigma) = -{\sqrt{3}\over 2}N.\label{e22}
\end{eqnarray}

The numerical values of Eqs.(20)--(22) coincide almost identically with the
results of the $SU(6)$--based NRQM taking account of the $SU(3)$ breaking
due to the quark--mass differences \cite{a_8}. We stress, however, that no
NR assumption or explicit $SU(6)$-wave function are used this time.
The ratio $\alpha$ ( cf. Eq.(7) ) equals .61 in this case,
demonstrating a substantial influence of the nonvalence degrees of freedom
on this important parameter.

In calculations we used baryon magnetic moments from the PDG--tabulation
\cite{a_1} and new values of the $\mu(\Xi^-)$ and
$\mu(\Sigma^+)$, from \cite{a_2} and \cite{a_3}, respectively.

\section {\bf The OZI-rule violation in magnetic and axial couplings
of baryons}

Here, we follow a complementary view of the
nucleon structure, absorbing C(N's) into products of the corresponding
$\mu(q)$ and g(N's), keeping the constraint u/d=--2,and $\delta(B's)$
non-zero.  We shall refer to this approach \cite{a_10} as a correlated current
quark picture of nucleons. Then, instead of Eq.(11) we have (in n.m.)
\begin{eqnarray} && \Delta(N) = {1\over 6}(3(P+N) - \Sigma^+  + \Sigma^-
-\Xi^0 +\Xi^-)= -.07\pm.01,\label{23}\\ && \mu_N(\overline ss)=\mu(s) \langle
N|\overline ss|N\rangle= (1- {d\over s})^{-1}\Delta =.13,\label{24}
\end{eqnarray}
where independence of the sum P+N of the C(N's) and the ratio d/s=1.55
from the correspondingly modified Eq.(12) were used.
By definition, $\mu_N(\overline ss)$ represents the contribution of
strange ("current") quarks to nucleon magnetic moments.Numerically, Eq.(24)
agrees fairly well with other more specific models ( see,e.g. \cite{a_11}).
The calculated quantity indicates violation of the OZI rule and the strange
current quark contribution $\mu_N(\overline ss)$ is seen to constitute
a sizable part of the isoscalar magnetic moment of nucleons (or, which is
approximately the same, of the nonstrange constituent quarks)
\begin{eqnarray}
{1\over 2}(P+N)=\mu(\overline uu + \overline dd)
+\mu(\overline ss) =.44,\label{25}
\end{eqnarray}
This observation helps to understand the unexpectedly
large ratio \cite{a_12}
\begin{eqnarray}
BR\left ( {{\overline PN\to \phi +\pi}
\over {\overline PN\to \omega +\pi}}\right )\simeq(10\pm2)\% ,\label{e26}
\end{eqnarray}
reported for the s--wave $\bar NN$ -- annihilation reaction.

Indeed, the transition ${(\overline PN)}_{s-wave} \to V+\pi$,
where $V=\gamma,\omega,\phi$, is of the magnetic dipole type. Therefore,
the transition operator should be proportional to the isoscalar
magnetic moment contributions
from the light u-- and d--quarks and
the strange s--quark, Eq.(25). The transition
operators for the $\omega$-- and $\phi$--mesons are obtained from
$\mu(\overline qq)$ and $\mu(\overline ss)$  through the well--known
vector meson dominance model ( VDM ).Using  the "ideal" mixing ratio
$g_\omega:g_\phi=1:\sqrt 2$ for the
photon--vector--meson junction couplings
and $\mu_\omega:\mu_\phi=\mu(\overline qq):\mu(\overline ss)$
according to Eq.(24) and Eq.(25), we get
\begin{eqnarray}
BR\left ({{ \overline PN \to \phi +\pi} \over
{\overline PN \to \omega +\pi}}\right )
\simeq \left ({ \mu( \overline ss)} \over
{\sqrt 2 \mu( \overline uu + \overline dd)}
\right )^2 \left ({p_\phi ^{c.m.}}
\over {p_\omega^{c.m.}} \right )^3 \simeq  6\%,\label{e27}
\end{eqnarray}
which is reasonably compared with data.

The structure constants connected with the vector part of the weak neutral
current ( NC ) are obtained from the
electromagnetic ones by the substitution
\begin{eqnarray}
Q(q)\to V^{NC}(q) =\frac{1}{2sin\theta_W cos\theta_W }(t_L(q)-2Q(q)\sin^2\theta_W),\label{e28}
\end{eqnarray}
where Q(q) is the quark electric charge,
$t_L$--the 3rd component of the weak isospin,
$ sin^2\theta_W=.23 $, $ \theta_W $--the weak angle. In this way,
for the NC analogue of magnetic moments of the proton, neutron and
deuteron we get ( in the units of n.m.)
\begin{eqnarray}
&& \mu^{NC}(P) \simeq 1.49 (1.25),\label{e29}\\
&& \mu^{NC}(N) \simeq -1.52 (-1.73),\label{e30}\\
&& \mu^{NC}(d) \simeq -0.04 (-.47),\label{e31}
\end{eqnarray}
where the values in the parentheses
refer to the neglect of the strange quark contributions.
Therefore the planned detailed investigation of the $\gamma-Z^0$
interference effects and measuring $\mu_N(\overline ss)$ via the P- odd
effects in polarized electron--nucleon scattering \cite{a_11} will give
an important information on the strangeness content of the nucleon.

The value of $ \langle N|\overline ss|N\rangle $
in Eq.(25) can be considered as the difference of the averaged values
$ \langle N|{l_z(s)} + {\sigma_z(s)}|N\rangle $
of strange quarks and antiquarks, respectively,
$ l_z(s) $ and $ \sigma_z(s) $ being the orbital and spin
operators of corresponding
quarks in the polarized nucleon. The product of this combination of
the quark angular momenta and the effective magnetic moments of quarks,
in which quark energies are used instead of
quark masses, was proposed \cite{a_13} to define
the magnetic moment operator in a relativistic model of quark composites.
Taking, for the sake of qualitative estimates,
$ \mu(s) $ =$ -{e/(6 \varepsilon_s)} $ $ \simeq -.7 $ n.m., where we put
$ {\varepsilon_s} \simeq {({m_s}^2 + {p_s}^2)^{1 \over 2}} $,
$ {m_s}(\simeq 150 MeV) $ -- the "current" quark mass,
$ {p_s}(\simeq 400 MeV) $--the mean momentum of sea quarks,
we obtain
\begin{eqnarray}
\sum_{s,\bar s}\langle N,J_{z}=
+{1\over 2}| \sigma_{z}(s)
+ l_{z}(s) - \sigma_{z}(\bar s) - l_{z}(\bar s)|N,J_{z}=
+{1\over 2}\rangle\simeq  -.17, \label{e32}
\end{eqnarray}
which can be compared with the other strange quark spin characteristics
$ {\Delta s} \simeq -.1 $, as given by the polarized DIS
data, e.g. \cite{a_14}.
We wish here to note the following alternative. As is known \cite{a_14},
to obtain the contributions of the u-,d-,
and s-flavoured quarks to the proton
spin, denoted by $\Delta u(p), \Delta d(p)$ and $\Delta s(p)$, the use is
usually made of baryon semileptonic weak decays
treated with the help of the exact $SU(3)$-symmetry. It has been shown
earlier \cite {a_15,a_16}
that when both the strangeness-changing
$(\Delta S=1)$ and strangeness-conserving
$(\Delta S=0)$ transitions are taken for the analysis,
then $(D/F+D)_{ax}^{\Delta S=0,1} = .635 \pm .005 $ while
$(D/F+D)_{ax}^{\Delta S=0}=.584 \pm .035 $, which is close to
$ (D/D+F)_{mag} \simeq .58 $, according to Eq.(7).
We list below two sets of the $\Delta q$-values,
we have obtained from the data
with inclusion of the QCD radiative corrections
(e.g.\cite{a_14} and references
therein) : $\Delta u(p) \simeq .82 (.83),\;
\Delta d(p) \simeq -.44(-.37),\;
\Delta s =-.10 \pm .04(-.19 \pm .05)$, where the values in
the parentheses correspond to $\alpha _{D}=(D /D+F)=.58$.
At the same time, the problem of
difference of the following two expressions
\begin{eqnarray}
&& F - D = \Delta d(p) - \Delta s(p) =
G_A^{exp}(\Sigma^{-}\rightarrow n) = -.34 \pm .02,\label{e33)}\\
&& F - D = \Delta d(p) - \Delta s(p) =
G_A^{exp}(n \rightarrow p) - \sqrt6
G_A^{exp}(\Sigma \rightarrow \Lambda)= -.19 \pm .04,\label{e34}
\end{eqnarray}
of which we prefer the second one
when we postulate $\alpha_{D} ^{ax} =\alpha _{D}^{mag}$,
remains largely open.The intriguing possibility can, however, be mentioned
that the numerical value of the $G_A^{exp}(\Sigma^{-}\rightarrow n)$,
coinciding with Eq.(34), was in fact found in \cite{a_17}, if the weak--electric
dipole form factor, referred as one of the second class current effects,
is included in the joint analysis of all experimental data.\\
We note in passing, that the production of the axial -- vector meson
composed of strange quark-antiquark pairs in antinucleon annihilation may
also be useful to
probe the strange content of the nucleon and the dynamics of the OZI rule
violation.The $ q + (\bar s s)$- configuration with $ J^P= {1 \over 2}^{+}$
composing in part the nonstrange constituent quark can evolve via a chain of
transitions
\begin{equation}
q + (\bar ss)_{vac} \rightarrow (\bar sq)_{0^{-}}+ s \rightarrow q + (\bar ss)_{J^{PC}}
\end{equation}
with $J^{PC}= 0^{++},1^{++},1^{+-}$.A simple recoupling of the angular
momenta enable, through the pertinent Clebsch--Gordan coefficients,
to obtain then the ratio
\begin{equation}
w(1^{++}) : w(1^{+-}) : w(0^{++}) = 2 : 1 : 1
\end{equation}
which may serve as a qualitative measure of relative yields of corresponding
meson states, composed mainly of the "strange matter".The relevant mesons
with $ J^P=1^{+}$ might be the $f_1(1420;1^{++})$-- or $f_1(1510;1^{++})$--
and $h_1(1380;1^{+-(?)})$--resonances \cite{a_1}.

\section {\bf Remarks}

1. The deviation of the ratio
$F/D=.75$, Eq.7, from the $SU(6)$ --value $2/3$
shows, that despite the validity of the celebrated
$SU(6)$--ratio \cite{a_9} $\mu(P)/\mu(N)=-3/2$, the $SU(6)$--symmetry
is strongly broken. The importance of taking into account the nonvalence
degrees of freedom in relevant parametrization of the observables
within the (broken) internal symmetries is demonstrated

2. The real meaning of all failures ( and relative successes, of course )
of the "naive" NRQM results ( e.g.\cite{a_8}), coinciding with values from
Eqs.(20)--(22), is the neglect
of contributions due to the nonvalence degrees
of freedom ( the sea partons and/or meson clouds
at the periphery of hadrons).

3. The strange current quarks considered as
a part of the constituent quark
internal structure should be explored by the probes with the resolution
capability comparable with the constituent quark
size, i.e. in the processes
with high enough momentum transfers or the energy release.The OZI--rule
violating $\phi$ - meson production in
the proton--antiproton annihilation
gives further evidence of the polarized strange
quark sea inside the polarized
nucleon.\\

The author is grateful to M.G.Sapozhnikov for a useful discussion of the data
on the OZI -- rule violation in the antinucleon annihilation reactions.
This work was supported in part by the Russian Foundation for Basic
Researches (grant No.96-02-18137).

\end{document}